\documentclass[twocolumn,twoside,prl,superscriptaddress]{revtex4}
\usepackage{amsmath}
\usepackage{amssymb}
\usepackage{epsfig}

\newcommand{\ket}[1]{| #1 \rangle}
\newcommand{\bra}[1]{\langle #1 |}
\newcommand{\proj}[1]{\ket{#1}\!\bra{#1}}
\newcommand{\braket}[2]{\left\langle #1| #2 \right\rangle}
\newcommand{\op}[2]{ | #1 \rangle \! \langle #2 |}
\newcommand{\tr}{\operatorname{tr}}
\newcommand{\av}[1]{\left\langle #1 \right\rangle_t}
\newcommand{\eff}{\rm eff}

\newcommand{\mM}{\mathcal{M}}

\newcommand{\mH}{\mathcal{H}}
\newcommand{\ident}{I}

\usepackage{amsthm}
\newtheorem{theorem}{Theorem}
\newtheorem{corollary}{Corollary}

\begin{document}

\title{Equilibration of quantum systems and subsystems}

\author{Anthony J. Short}\email{a.j.short@damtp.cam.ac.uk}
\affiliation{Department of Applied Mathematics and Theoretical Physics, University of Cambridge, Centre for
Mathematical Sciences, Wilberforce Road, Cambridge CB3 0WA, U.K.}

\begin{abstract}
We unify two recent results concerning equilibration in quantum theory. We first generalise a proof of Reimann [PRL 101,190403 (2008)], that the expectation value of `realistic' quantum observables will equilibrate under very general conditions, and discuss its implications for the equilibration of quantum systems. We then use this to re-derive an independent result of Linden et. al. [PRE 79, 061103 (2009)], showing that small subsystems generically evolve to an approximately static equilibrium state. Finally, we consider subspaces in which all initial states effectively equilibrate to the same state.
\end{abstract}

\maketitle

\section{Introduction}

Recently there has been significant progress in understanding the foundations of statistical mechanics, based on fundamentally quantum arguments~\cite{Mahler, Goldstein1, Goldstein2, PopescuShortWinter, Tasaki, reimann1, reimann2, us1,us2, gogolin1, gogolin2}. In particular, Reimann \cite{reimann1, reimann2} has shown that the expectation value of any `realistic' quantum observable will equilibrate to an approximately static value, given very weak assumptions about the Hamiltonian and initial state. Interestingly, the same assumptions were used independently by Linden \emph{et al.} \cite{us1, us2}, to prove that any small quantum subsystem will evolve to an approximately static equilibrium state (such that even `unrealistic' observables on the subsystem equilibrate). In this paper we unify these two results, by deriving the central result of Linden \emph{et al.}\cite{us1} from a generalisation of Reimann's result. We also offer a further discussion and extension of  Reimann's results, showing that systems will appear to equilibrate with respect to all reasonable experimental capabilities. Finally, we identify subspaces of initial states which equilibrate to the same state.

\section{Equilibration of expectation values.}

We  prove below a generalisation of Reimann's result  that the expectation value of any operator will almost always be close to that of the equilibrium state \cite{reimann1}. We extend his results to include non-Hermitian operators (which we will use later to prove equilibration of subsystems), correct a subtle mistake made in \cite{reimann2} when considering degenerate Hamiltonians, and improve the bound obtained  by a  factor of 4. As in \cite{reimann2, us2}, we make one assumption in the proof, which is that the Hamiltonian has \emph{non-degenerate energy gaps}.  This means that given any four energy eigenvalues $E_k, E_l, E_m$ and $E_n$,
\begin{equation}  \label{eq:non-degen}
E_k - E_l = E_m - E_n  \Rightarrow \begin{array}{c} (E_k = E_l \; \textrm{and}\; E_m = E_n) \\ \textrm{or}  \\ (E_k = E_m \; \textrm{and}\; E_l = E_n). \end{array}
\end{equation}
 Note that this definition allows degenerate energy levels, which may arise due to symmetries. However, it ensures that all subsystems physically interact with each other. In particular, given any decomposition of the system into two subsystems $\mH = \mH_A \otimes\mH_B$, equation  (\ref{eq:non-degen}) will not be satisfied by any Hamiltonian of the form $H=H_A \otimes I_B + I_A \otimes H_B$ (unless either $H_A$ or $H_B$ is proportional to the identity) \footnote{To see this, consider the four energy eigenstates $\ket{k}=\ket{0}\ket{0}$, $\ket{l}=\ket{0}\ket{1}, \ket{m}= \ket{1}\ket{0}, \ket{n}=\ket{1}\ket{1}$ which are products of eigenstates of $H_A$ and $H_B$.}.

\begin{theorem}[Generalisation of Reimann's result \cite{reimann1}] Consider a $d$-dimensional quantum system evolving under a Hamiltonian $H=\sum_n E_n P_n$, where $P_n$ is the projector onto the eigenspace with energy $E_n$. Denote the system's density operator by $\rho(t)$, and its time-averaged state by $\omega \equiv \av{\rho(t)}$. If $H$ has non-degenerate energy gaps, then for any operator $A$,
\begin{equation} \label{eq:theorem}
\sigma_A^2 \equiv \av{ \left| \tr \left(A \rho\left(t\right) \right) - \tr \left( A \omega\right) \right|^2 } \leq \frac{\Delta(A)^2 }{4 d_{\eff}} \leq \frac{\|A\|^2}{d_{\eff}}
\end{equation}
where $\|A\|$ is the standard operator norm \footnote{$ \|A\|=\sup \{ \sqrt{\bra{v} A^{\dag} A \ket{v}} : \ket{v} \in \mH \,\textrm{with}\, \braket{v}{v}=1\}$, or equivalently $\|A\|$ is the largest singular value of $A$.},
\begin{equation}
\Delta(A) \equiv 2 \min_{c \in \mathbb{C}} \| A- c I \|, 
\end{equation}
and
 \begin{equation}
 d_{\eff}  \equiv \frac{1}{\sum_n \big( \tr(P_n \rho(0)) \big)^2}.
\end{equation}
\end{theorem}

This bound will be most significant when the number of different energies incorporated in the state, characterised by the effective dimension $ d_{\eff}$, is very large. Note that $1 \leq d_{\eff} \leq d$, and that $d_{\eff}=N$ when a measurement of $H$ would yield $N$ different energies with equal probability. For pure states $d_{\eff} = \tr(\omega^2)^{-1}$ as in \cite{us1, us2}, but it may be smaller for mixed states when the Hamiltonian is degenerate.

The quantity $\Delta(A) $ gives the range of eigenvalues when $A$ is Hermitian, and gives a slightly tighter bound than the operator norm.  Following \cite{reimann2}, we could improve the bound further by replacing $\Delta(A) $ with a state- and Hamiltonian-dependent term \footnote{In particular we could replace $\Delta(A)$ with  $\Delta''(A)=\min_{\tilde{A}} 2\|\tilde{A}\|$, where the operators $\tilde{A}$ are obtained by subtracting any function of $H$ from $A$ and projecting onto the support of $\omega$.}, however we omit this step here for simplicity.

\textbf{Proof:} To avoid some difficulties which arise when considering degenerate Hamiltonians, we initially consider a pure state $\rho(t) = \proj{\psi(t)}$, then extend the results to mixed states via purification.

We can always choose an energy eigenbasis such that  $\ket{\psi(t)}$ has non-zero overlap with only a single energy eigenstate $\ket{n}$ of each distinct energy, by including  states $\ket{n} = P_n \ket{\psi(0)}/\sqrt{\bra{\psi(0)} P_n \ket{\psi(0)}}$ whenever $\bra{\psi(0)} P_n \ket{\psi(0)}>0$. The state at time $t$ is then given by
 \begin{equation}
\ket{\psi(t)} = \sum_{n} c_n e^{-i E_n t/\hbar} \ket{n},
\end{equation}
where $c_n = \braket{n}{\psi(0)}$. This state will evolve in the subspace spanned by $\{\ket{n}\}$ as if it were acted on by the non-degenerate Hamiltonian $H'=\sum_n E_n \proj{n}$. For any operator $A$, it follows that
\begin{eqnarray}
 \sigma_A^2\!\!\! &=&  \av{ |\tr(A [\rho(t) - \omega] )|^2 } \nonumber \\
 &=& \av{ \left|\sum_{n \neq m} c_n c_m^* e^{i(E_m-E_n)t/\hbar} \bra{m} A \ket{n} \right|^2} \nonumber \\
 &=& \!\!\!\! \sum_{\scriptsize \begin{array}{c} n \neq m \\ k\neq l \end{array}} \!\!\! \! c_n c_m^* c_k c_l^*\av{ e^{i(E_m-E_n + E_l - E_k)t/\hbar} }\bra{m} A \op{n}{l} A^{\dag} \ket{k} \nonumber \\
 &=& \sum_{n,m} |c_n|^2 |c_m|^2 \bra{m} A \op{n}{n} A^{\dag} \ket{m} - \sum_{n} |c_n|^4 |\bra{n}A \ket{n}|^2 \nonumber \\
 & \leq &\tr( A \omega A^{\dag} \omega )  \nonumber \\
 & \leq & \sqrt{\tr(A^{\dag}\!A\, \omega^2) \tr (A A^{\dag} \omega^2)} \nonumber \\
 &\leq& \| A \|^2 \tr(\omega^2)  \label{eq:pure_theorem} \nonumber \\
&=& \| A \|^2\tr\left[ \left(\sum_n  |c_n|^2  \proj{n}\right)^2\right] \nonumber \\
&=& \| A \|^2 \sum_n \big( \tr(P_n \rho(0)) \big)^2 \nonumber \\
&=& \frac{\| A \|^2}{d_{\eff}}. 
\end{eqnarray}
In the fourth line, we have used the assumption that the Hamiltonian has non-degenerate energy gaps, in the sixth line we have used the Cauchy-Schwartz inequality for operators with  scalar product $\tr(A^{\dag} B)$ and the cyclic symmetry of the trace, and in the seventh line we have used the fact that for positive operators $P$ and  $Q$, $\tr(PQ) \leq \|P\| \tr(Q) $. This gives the weaker bound in the theorem.

To obtain the tighter bound, we note that $\sigma_A$ is invariant if $A$ is replaced by  $\tilde{A} = A- c I$ for any complex $c$. Performing this substitution with $c$ chosen so as to minimize $\|\tilde{A}\|$ we can replace $\|A\|$ with $\|\tilde{A}\|=\Delta(A)/2$.

An extension to mixed states can be obtained via purification, following the approach discussed in \cite{us2}. Given any initial state $\rho(0)$ on $\mH$, we can always define a pure state $\ket{\phi(0)}$ on $\mH \otimes \mH$ such that the reduced state of the first system is $\rho(0)$. By evolving $\ket{\phi(t)}$ under the joint Hamiltonian $H'=H \otimes I$, we will recover the correct evolution  $\rho(t)$ of the first system, and $H'$ will have non-degenerate energy gaps whenever $H$ does.  The expectation value of any operator $A$ for $\rho(t)$ will  be the same as the expectation value of $A'=A \otimes I$ on the total system, and we also obtain $\Delta(A')=\Delta(A)$, $\|A\|=\|A'\|$, and $d_{\eff}'=d_{\eff}$. However, note that $\tr(\omega'^2)$ does not equal $\tr(\omega^2)$.  Using the result for pure states, we can obtain (\ref{eq:theorem})  in the mixed state case from
 \begin{equation}
 \sigma_A^2 = \sigma_{A'}^2 \leq  \frac{\Delta(A')^2 }{4 d_{\eff}'} =\frac{\Delta(A)^2 }{4 d_{\eff}}.
\end{equation}
This completes the proof. $\square$

In \cite{reimann1}, Reimann proves that $\sigma_A^2 \leq \Delta(A)^2 \tr(\omega^2)$ when $A$ is Hermitian and the Hamiltonian has non-degenerate levels as well as non-degenerate gaps. However, it appears that there is a subtle mistake in \cite{reimann2} when extending this proof to degenerate Hamiltonians. Specifically, the step from equation (D.11) to (D.12) in \cite{reimann2} does not follow if the state has support on more than one energy eigenstate in a degenerate subspace. A counterexample is provided by the mixed state $\rho(0) = \frac{1}{k} \proj{0} \otimes \ident$, of a qubit and a $k$-dimensional system, with $H=(\op{0}{1} + \op{1}{0}) \otimes \ident$ and $A = (\proj{0}-\proj{1}) \otimes \ident$. In this case $\sigma^2_A = \frac{1}{2}$, $\Delta(A)=2$ and $\tr(\omega^2) = \frac{1}{2k}$, giving $\sigma_A^2 > \Delta(A)^2 \tr(\omega^2)$ when $k>4$. However, subsequently in \cite{reimann2}, $ \tr(\omega^2)$ is replaced by an upper bound of $\max_n \tr(\rho(0) P_n)$, and this also upper bounds $d_{\eff}^{-1}$, so later results are unaffected. Note that the bound  given by Theorem 1 for the same example is satisfied tightly for all $k$, as $ d_{\eff}=2$ and thus $\sigma_A^2 = \frac{1}{2} = \frac{\Delta(A)^2}{4 d_{\eff}}$.

\section{Distinguishability} When $A$ represents a physical observable and $\rho(0)$ a realistic initial state, it is argued in \cite{reimann1}   that the difference between $\tr (A \rho(t))$ and $\tr (A \omega)$ will almost always be less than realistic experimental precision. This is then taken to imply that $\rho(t)$ will be indistinguishable from $\omega$ for the overwhelming majority of times.

However, the fact that two states yield the same expectation value for a measurement does not necessarily imply that they cannot be distinguished by it. For example, a measurement yielding an equal mixture of  $+1$ and $-1$ outcomes for one state and always yielding $0$ for a second state clearly can distinguish the two states, despite the expectation values in the two cases being identical. Furthermore, even though any particular realistic  measurement cannot distinguish $\rho(t)$ from $\omega$ for almost all times, this does not imply that for almost all times, no realistic measurement can distinguish $\rho(t)$ from $\omega$. This is because the optimal measurement to distinguish the two states may change over time. Finally, the measurement precision is not easy to define for measurements with discrete outcomes.

To address these issues, we first note that the most general quantum measurement is not described by a Hermitian operator, but by a positive operator valued measure (POVM). For simplicity, we consider POVMs with a finite set of outcomes, which is reasonable for realistic measurements, as even continuous outputs such as pointer position cannot be determined or recorded with infinite precision \footnote{However, our results could be extended to continuous output sets using measure theory if desired}. A general measurement $M$ is described by giving a positive operator $M_r$ for each possible measurement result  $r$, satisfying $\sum_r M_r= I$. The probability of obtaining  result $r$ when measuring $M$ on $\rho$ is given by $\tr(M_r \rho)$.

Suppose  you are given an unknown quantum state, which is either $\rho_1$ or $\rho_2$ with equal probability. Your maximum success probability in guessing which state you were given after performing the measurement $M$ is 
\begin{equation}
p^{\textrm{succ}}_{M} = \frac{1}{2} ( 1 + D_{M} (\rho_1, \rho_2))
\end{equation}
where
\begin{equation}
 D_{M} (\rho_1, \rho_2) \equiv \frac{1}{2} \sum_{r} | \tr(M_r \rho_1) - \tr(M_r \rho_2) |.
\end{equation}
We refer to $D_{M} (\rho_1, \rho_2)$ as the distinguishability of $\rho_1$ and $\rho_2$ using the measurement $M$. 
Similarly, the distinguishability of two states using any measurement from a set $\mM$  is given by
\begin{equation}
D_{\mM} (\rho_1, \rho_2) \equiv \max_{M \in \mM}   D_{M} (\rho_1, \rho_2).
\end{equation}
Note that
\begin{equation}
0 \leq D_{\mM} (\rho_1, \rho_2)  \leq D(\rho_1, \rho_2)  \leq 1,
\end{equation}
where $D(\rho_1, \rho_2) = \frac{1}{2} \tr|\rho_1 - \rho_2|$ is the trace-distance, which is equal to  $D_{\mM} (\rho_1, \rho_2)$ when $\mathcal{M}$ includes all measurements.

\section{Effective equilibration of large systems} For typical macroscopic systems, the dimension of $\mH$ will be incredibly large (e.g. For Avagardo's number $N_A$ of spin-$\frac{1}{2}$ particles, we would have $d >10^{10^{23}}$), and it is unrealistic to be able to perform any measurement with this many outcomes, let alone all such measurements.
For practical purposes, we are therefore restricted to some set of realistic physical measurements $\mM$. In this case, we would expect $\mM$ to be a finite set, as all realistic experimental setups (including all settings of variable parameters) will be describable within a finite number of pages of text.

We say that a state \emph{effectively equilibrates} if
\begin{equation}
\av{D_{\mM} (\rho(t), \omega)} \ll 1.
\end{equation}
This means that for almost all times, it is almost impossible to distinguish the true state $\rho(t)$ from the equilibrium state $\omega$ using any achievable measurement.

We can obtain an upper bound on the average distinguishability as a corollary of theorem 1. 

\begin{corollary}
Consider a quantum system evolving under a Hamiltonian with non-degenerate energy gaps. The average distinguishability of the system's state $\rho(t)$ from $\omega$, given a finite set of measurements $\mM$, satisfies
\begin{equation}
\av{D_{\mM} (\rho(t), \omega)} \leq  \frac{\sum_{M \in \mM} \sum_{r  } \Delta(M_r)}{4\sqrt{d_{\eff}}} \leq \frac{N(\mM)}{4\sqrt{d_{\eff}}}, \label{eq:effective_equilibration}
\end{equation}
where $N(\mM)$ is the total number of  outcomes for all measurements in $\mM$.
\end{corollary}
The first bound will be tighter when measurements are imprecise, as each outcome is weighted by $\Delta(M_r) \in [0,1]$, reflecting its usefulness in distinguishing states \footnote{Note that $\Delta(M_r)$ is the maximum difference in probability of that result occurring for any two states}.

\textbf{Proof:}
\begin{eqnarray}
\av{D_{\mM} (\rho(t), \omega)}  &=&  \av{ \max_{M(t) \in \mM} D_{M(t)} (\rho(t), \omega)} \nonumber\\
&\leq&  \sum_{M \in \mM} \av{ D_M (\rho(t), \omega)} \nonumber\\
&=&\frac{1}{2}  \sum_{M \in \mM}  \sum_{r  } \av{ | \tr(M_r \rho(t) )  - \tr(M_r \omega) | } \nonumber\\
&\leq&\frac{1}{2}  \sum_{M \in \mM} \sum_{r  } \sqrt{ \sigma_{M_r}^2  } \nonumber\\
&\leq & \frac{ \sum_{M \in \mM} \sum_{r  } \Delta(M_r)}{4\sqrt{d_{\eff}}} \nonumber\\
&\leq & \frac{N(\mM)}{4\sqrt{d_{\eff}}}. \label{eq:distinguishability} \qquad \square
\end{eqnarray}

In realistic experiments, we would expect the bound on the right of (\ref{eq:effective_equilibration}) to be much smaller than 1, implying that the state of the system effectively equilibrates to $\omega$. Consider again our system of $N_A$ spins. If $d_{\eff}\geq~d^{\,0.1} $, even if we take $\mM$ to include any experiment whose description could be written in $10^{19}$ words, each of which generates up to  $10^{21}$ bytes of data, we would still obtain $\av{D_{\mM} (\rho(t), \omega)} \leq 1/(10^{10^{22}})$.

 \section{Equilibration of small subsystems}

 Now consider that the system can be decomposed into two parts, a small subsystem of interest $S$, and the remainder of the system which we refer to as the bath $B$.  Then $\mH = \mH_S \otimes \mH_B$, where $\mH_{S/B}$ has dimension $d_{S/B}$. It is helpful to define the reduced states of the subsystem $\rho_S(t) = \tr_B(\rho(t))$ and $\omega_S = \tr_B(\omega)$.

In such cases, it was shown in \cite{us1, us2} that for sufficiently large $d_{\eff}$ the subsystem's state fully equilibrates, such that for almost all times, no measurement  on the subsystem (even `unrealistic' ones) can distinguish $\rho(t)$ from $\omega$. In particular, when $\rho(t)$ is pure and the Hamiltonian has non-degenerate energy levels as well as non-degenerate energy gaps, it is proven in \cite{us1} that
\begin{equation} \label{eq:oureqn}
\av{D(\rho_S(t), \omega_S)} \leq \frac{1}{2} \sqrt{\frac{d_S^2}{d_{\eff}}}.
\end{equation}
Extending this result to degenerate Hamiltonians and initially mixed states is discussed in \cite{us2}.

We cannot recover this bound directly from (\ref{eq:effective_equilibration}) by considering the set of all measurements on the subsystem, because this set contains an infinite number of measurements. However, we can derive (\ref{eq:oureqn}) from Theorem 1 by considering an orthonormal operator basis for the subsystem, given by the $d_S^2$ operators \cite{schwinger}
\begin{equation}
F_{(d_Sk_0 + k_1)} = \frac{1}{\sqrt{d_S}} \sum_{l} e^{\frac{2 \pi i l k_0}{d_S}} \ket{(l+k_1)\, \textrm{mod}\, d_S} \bra{l}
\end{equation}
where $k_0,k_1 \in \{0,1,\ldots d_S-1 \}$  and the  states $\ket{l}$ are an arbitrary orthonormal basis for the subsystem. Then writing $(\rho_S(t) - \omega_S) = \sum_{k} \lambda_k(t) F_k$ we have
\begin{eqnarray}
\av{D(\rho_S(t), \omega_S)}\!\! &=&\! \frac{1}{2} \av{\tr \big|\sum_k \lambda_k(t) F_k \big|}  \nonumber \\
&\leq&\! \frac{1}{2} \av{ \sqrt{ d_S  \tr \big(\sum_{kl} \lambda_k(t)  \lambda^*_l(t) F_l^{\dag} F_k \big)}}  \nonumber \\
&\leq&\! \frac{1}{2}  \sqrt{ d_S \sum_{kl}  \av{\lambda_k(t) \lambda^*_l(t) } \tr (F_l^{\dag} F_k)}  \nonumber \\
&=&\! \frac{1}{2} \sqrt{ d_S \sum_k \av{|\lambda_k(t)|^2}} \nonumber \\
&=&\! \frac{1}{2} \sqrt{ d_S \sum_k \av{\big|\tr\big((\rho(t) - \omega)F^{\dag}_k\! \otimes I  \big)\big|^2}} \nonumber \\
&\leq&\! \frac{1}{2} \sqrt{ d_S \sum_k \frac{\|F^{\dag}_k\! \otimes I\|^2 }{ d_{\eff}}} \nonumber \\
&\leq&\! \frac{1}{2} \sqrt{ \frac{d_S^2}{d_{\eff}}}.
\end{eqnarray}
In the second line we have used a standard relation between the 1- and 2-norm, and in the sixth line we have used Theorem 1 for the non-Hermitian operator $F^{\dag}_k\! \otimes I$. Note that $\sqrt{d_S} F_k$ is unitary, and thus $\|F_k^{\dag} \otimes I\| = \frac{1}{\sqrt{d_S}}$.

\section{Universality of equilibrium states} We have so far been concerned with when states equilibrate, rather than the nature of their equilibrium state. However, one of the notable properties of equilibration is that many initial states effectively equilibrate to the same state, determined only by macroscopic properties such as temperature. Given a particular Hamiltonian and a set of realistic measurements $\mM$, we can  construct a partition of the Hilbert space into a direct sum of subspaces $\mH = \bigoplus_k \mH_k$, such that all states within $\mH_k$ with large enough $d_{\eff}$ effectively equilibrate to the same state $\Omega_k$.

One way to  achieve this is to choose the subspaces such that each projector $\Pi_k$ onto $\mH_k$ commutes with the Hamiltonian, and such that any two energy eigenstates in $\mH_k$ are hard to distinguish. i.e. For some fixed $\epsilon$ satisfying $0 <  \epsilon \ll 1$, and all normalised energy eigenstates $\ket{i}, \ket{j} \in \mH_k$ 
\begin{equation}
 D_{\mM} (\proj{i}, \proj{j}) \leq \epsilon.
\end{equation}
When  $d_{\eff}$ is sufficiently large, it follows that all states in  $\mH_k$ effectively equilibrate to $\Omega_k = \Pi_k/ \tr(\Pi_k)$, as
\begin{eqnarray}
\av{D_{\mM}(\rho(t), \Omega_k)}\! &\leq& \av{D_{\mM}(\rho(t), \omega)} + \av{D_{\mM}(\omega, \Omega_k)} \nonumber \\
\! &\leq&\!\! \frac{N(\mM)}{4\sqrt{d_{\eff}}} + \sum_{i,j} \frac{\bra{i} \omega \ket{i}}{\tr(\Pi_k)} D_{\mM}(\proj{i}, \proj{j}) \nonumber \\
\! &\leq& \!\!\frac{N(\mM)}{4\sqrt{d_{\eff}}} + \epsilon.
\end{eqnarray}
where the sums in the second line are over an eigenbasis of $\omega$ (which is also a basis of  $\mH_k$), and we have used the fact that $D_{\mM} (\rho, \sigma)$ satisfies the triangle inequality ($D_{\mM} (\rho, \sigma) \leq D_{\mM} (\rho, \tau) + D_{\mM} (\tau, \sigma)$) and convexity,
\begin{equation}
D_{\mM}\left( \sum_i p_i \rho_i, \sigma \right) \leq  \sum_i p_i D_{\mM}(\rho_i, \sigma),
\end{equation}
where $p_i \geq 0$ and $\sum_i p_i= 1$. 

When $\mH_k$ can be chosen to be a small band of energies, the equilibrium state $\Omega_k$ will be the usual microcanonical state.

\section{Conclusions} To summarise, we have shown that two key results of \cite{us1, reimann1} about the equilibration of large systems can be derived from very weak assumptions  (non-degenerate energy gaps, and sufficiently large $d_{\eff}$), and a single theorem (Theorem 1). In particular, for almost all times, the state of an isolated quantum system will be indistinguishable from its equilibrium state $\omega$ using any \emph{ realistic} experiment, and the state of a small subsystem will be indistinguishable from $\omega_S$ using any experiment. 

Although the first result has a similar flavour to the classical equilibration of course-grained observables such as density and pressure, it is really much stronger, as it encompasses any measurement you could describe and record the data from in a reasonable length of text, including microscopic measurements. The second result has no classical analogue, as it yields an essentially static description of the true micro-state of a subsystem, rather than the rapidly fluctuating dynamical equilibrium of particles in classical statistical mechanics.  Given the difficulty of proving similar results in the classical case, it seems that  quantum theory offers a firmer foundation for statistical mechanics.

\medskip
\noindent
\textit{Acknowledgments.}
The author is supported by the Royal Society.

\end{document}